\documentstyle[prl,tighten,aps]{revtex}
\def\beqa{\begin{eqnarray}}
\def\eeqa{\end{eqnarray}}
\def\beq{\begin{equation}}
\def\eeq{\end{equation}}

\def\bib#1{$^{\ref{#1}}$}

\def\pr{{\it Phys. Rev.}\ }
\def\prl{{\it Phys. Rev. Lett.}\ }

\def\mnras{Mon. Not. R. Ast. Soc.\ }

\def\ie{{\it i.e. }}
\def\eg{{\it e.g. }}

\def\a{\`a }

\begin{document}
\def\bib#1{[{\ref{#1}}]}
   \title{\normalsize \bf Stability and Size of Galaxies from Planck's Constant}

\author{\normalsize 
Salvatore Capozziello$^{a,c,}$\thanks{E-mail:
capozziello@vaxsa.csied.unisa.it}, 
Salvatore De Martino$^{b,c,}$\thanks{E-mail:
demartino@physics.unisa.it},
Silvio De Siena$^{b,c,}$\thanks{E-mail:
desiena@physics.unisa.it}, and
Fabrizio Illuminati$^{b,c,}$\thanks{E-mail:
fabrizio@leopardi.phys.unisa.it} \\
{\small\em $^a$Dipartimento di Scienze Fisiche ``E. R. Caianiello",}\\ 
{\small\em $^b$Dipartimento di Fisica,}\\ 
{\small\em $^c$INFN, Sez. di Napoli and INFM, Unit\a di Salerno, }\\
{\small\em Universit\`a di Salerno, I-84081 Baronissi (SA), Italy.}} 
\date{\today}
\maketitle

	      \begin{abstract}
Stability and characterisitic geometrical and kinematical sizes of galaxies
are strictly related to a minimal characteristic action whose value is of
order $h$, the Planck constant. We infer that quantum mechanics, in some 
sense, determines the structure and the size of galaxies.
               \end{abstract}

\vspace{20. mm}

\vspace{20. mm}

\noindent  From a classical point of view, there are no arguments capable
of completely explaining the stability and the size of galaxies: the only
assumption is that they are considered to be 
relaxed (and virialized) systems where 
gravity is the overall interaction \bib{binney}. Such a force is considered 
as a ``Newtonian" interaction and the confining potentials, due to the mutual
attractions of stars (and the other components as dust and gas clouds)
can have several forms. For example, logarithmic potentials well describe the
regular motion of stars before the onset of chaos \bib{saasfee}.
In any case, the ``stability is an assumption" and sizes are deduced from 
observations.
Actually, the problem is extremely involved since galaxies undergo 
environmental effects, being never isolated systems; they always belong to 
large gravitationally bound systems as loose and tight groups, associations
or clusters of galaxies and the observational times are so short that the 
overall dynamics can be only inferred \bib{binney},\bib{vorontsov}.
Besides, galaxies have to be related to some cosmological model and, due to 
cosmological evolution of large scale structures, they should be connected to
some theory of primordial perturbations \bib{kolb},\bib{sakharov}.  
For these reasons, it is
not senseless to ask for some quantum signature in the today observed 
galaxies \bib{capozziello}.
The main point, however, is to connect the estremely large size of galaxies
($\sim 10$kpc) with the extremely small numbers of quantum mechanics
($h\sim 10^{-27}$ erg sec).

In this letter, we want to show that, for a given galaxy, a minimal 
characteristic action is of the order of $h$ and, furthermore, the onset of
chaos \bib{lyndenbell} 
is prevented if and only if the characteristic sizes of a galaxy are 
related to the Planck constant.
In other words, it is the quantum signature which stabilizes the galaxies;
furthermore it gives rise to their characteristic sizes, where, 
by ``sizes", we intend 
geometrical and kinematical quantities which assignes a 
galaxy. 

This result is not  particular since the collective features
(in particular the stability and the confinement properties) of several 
mesoscopic and macroscopic systems can be explained only by invoking quantum 
coherence on large scales (see for example \bib{galgani},\bib{bocchieri}).
Very famous examples of this new trend in physics are the high $T_{c}$ 
superconductivity systems or the optical fibres. Besides, it is possible to 
show that $h$ is the characteristic action for several macroscopic systems.
The scheme is: given a classical law of force $F(R)$, describing a system 
where $N$ particles are interacting on a length scale $R$, a characteristic
action of order $h$ is recovered. 

The ``classical" force $F(R)$ can be the electromagnetic interaction of
accelerator beams, the strong interaction of quark aggregates 
\bib{demartino}, or the Newtonian interaction acting on all the nucleons 
which are present in the Universe \bib{calogero}. Taking into account 
generalized theories of gravity \bib{stelle},\bib{brans},\bib{bertolami}
 which give 
corrective terms to the Newtonian potential in the weak energy limit, it is 
possible to show that any gravitationally bound system, where gravity is the
only overall interaction, undergoes this scheme \bib{capozziello}.

A heuristic argument can be given considering the total action for a bound, 
virialized system where $N$ is the number  of  constituents.
Let $E$ be the total energy so that the system is bound and stable. 
Let ${\cal T}$ be the characteristic time of the system 
(\eg the time in which a particle crosses the system, or the time in which
the system evolves and becomes relaxed). Combining these two quantities,
we get
\beq
\label{1}
{\cal A}\cong E{\cal T}\,,
\eeq
which is the total action. The only hypothesis which we need is that the 
system could undergo a time--statistical fluctuation, so that the 
characteristic time $\tau$ for the stochastic motion per particle will be
\bib{demartino},\bib{calogero} 

\beq
\label{2}
{\tau}\cong\frac{{\cal T}}{\sqrt{N}}\,.
\eeq
This hypothesis naturally emerges from the fact that a galaxy can be
treated as a statistical system \bib{binney}.

Immediately, we can define an energy per particle
\beq
\label{3}
\epsilon\cong\frac{E}{N}\,,
\eeq
and then a characteristic unit of action per particle is
\beq
\label{4}
\alpha=\epsilon\tau\cong\frac{{\cal A}}{\sqrt{N^3}}\,.
\eeq
These formal considerations can be applied to bound physical systems where 
the degrees of freedom have acquired the same energy (that is are 
virialized). It can be shown that for several
systems (among them also the whole observable Universe), it is
\beq
\label{5}
\alpha\simeq h\,,
\eeq
with an error of approximatively an order of magnitude
\bib{demartino},\bib{calogero}.

Let us now consider galaxies. The onset of chaos \bib{contopoulos},
in a realistic galactic potential, is for an energy per unit of mass
of the order $10^{15}$ (cm/sec$)^2$ while the period of a galactic rotation,
which can be assumed as a characteristic time, is about
\beq
\label{7}
{\cal T}_{rot}=3\times 10^{15}\mbox{sec}\,.
\eeq
The total mass of a typical galaxy is
\beq
\label{8}
M\cong 2\times 10^{44}\mbox{gr}\,.
\eeq
From Eq.(\ref{1}), combining these numbers, we get
\beq
\label{9}
{\cal A}\cong 10^{74}\mbox{erg sec}\,.
\eeq 
The number of nucleons in a star of a solar mass is $\sim 10^{57}$ and then,
for a galaxy,
\beq
\label{10}
N\cong 10^{68}\,.
\eeq
Introducing these numbers inside Eq.(\ref{4}), we get, with an error of an
order of magnitude, that 
{\it the characteristic unit of action for a galaxy is of the order of
Planck constant}. It is interesting to stress the fact that also if 
dark matter is considered into dynamics the result does not change
dramatically since the mass to luminosity ratio is of the order
$10\div 100$.

It is interesting to note that the values which we have used are on the 
boundary for the onset of chaos and the galaxy is assumed stable.
In other words, the stability of the system is related to the quantum 
mechanics.
Furthermore, the stability and the connection to quantum mechanics scale with
the number of particles. 

More formally, the characteristic unit of action can be derived for a system 
where a classical law of force $F(R)$ acts on the constituents of mass $m$
over a global size $R$. If the system is stable and virialized, the 
characteristic work done by the system is
\beq
\label{25}
{\cal L}\cong mv^{2}\,,
\eeq
and then
\beq
\label{26}
{\cal L}\cong NF(R)R\,.
\eeq
Using Eqs.(\ref{2}), (\ref{3}) and (\ref{4}), one gets
\beq
\label{27}
\alpha\cong m^{1/2}R^{3/2}\sqrt{F(R)}\,,
\eeq
indipendently of the type of force. In all cases one obtains
$\alpha\cong h$
\bib{galgani},\bib{demartino}. In particular, the result holds for 
gravitationally bound systems which can be globular clusters, galaxies, and 
clusters of galaxies \bib{capozziello}, up to the 
whole universe \bib{calogero}.
In these cosmological cases, the gravitational coupling, 
\ie the Newton coupling
$G_{N}$ must scale with the distance as several modified quantum theories of 
gravity imply \bib{stelle},\bib{brans},\bib{fradkin},\bib{sanders}.
However, the modification of $G_{N}$ is small and Newtonian gravity holds in 
the weak energy limit. 
Confirmations of this scheme are coming from satellites' measurements of
 long range acceleration \bib{pioneer}.
Gravitational potentials like
\beq
\label{28}
V(R)=-\frac{G(R)M}{R}\,,
\eeq
with
\beq
\label{29}
G(R)=\chi G_{N}\left(\frac{R}{R_{0}}\right)^{\eta}
\ln\left(\frac{R}{R_{0}}\right)\,,
\eeq
or
\beq
\label{30}
G(R)=G_{N}\left[1+a_{0}\exp(-R/R_{0}]\right)\,,
\eeq
well describe this situation. The parameters $\chi,\eta,a_{0}$ depend on the 
modified theory of gravity used \bib{bertolami},\bib{avramidy}.
$R_{0}$ can be assumed, for galaxies, of the order $\simeq 10$kpc \bib{sanders}.

An important point must be stressed. All these simple hypotheses
do not work for a single star which, from our point of view, is not 
properly a gravitationally bound system. In fact, nuclear and electromagnetic
interactions contribute to the stability of the system so that it cannot be
simply schematized only with a classical force acting on it.

This result, as we said above, holds also for other mesoscopic and 
macroscopic systems as  accelerator beams,  quark condensates and 
Bose--Einstein condensates \bib{demartino}. The rule, thus, 
seems general and it works also at astrophysical scales as 
those of galaxies. In all cases these situations,
involving complex aggregates which
exhibit a nontrivial interplay between mechanical or quantum mechanical
effects and thermodynamical and statistical effects, one can suitably 
define effective scales of length,  velocity, and  energy,
as well as  effective temperatures. We will now show how all these allow
to derive some interesting results on the
geometrical size of galaxies,  on the average thermal velocities
and wavelenghts for the galaxies.

Let us first introduce the ``emittance", which is a scale of length
(or, equivalently, of ``temperature") related to a given complex,
correlated system (such as a Bose condensate or a
charged particle beam in an accelerator \bib{demartino}). 
It can be defined as
\beq
\label{11}
{\cal E}\cong\lambda_{c}\sqrt{N}\,,
\eeq
where
\beq
\label{12}
\lambda_{c}=\frac{h}{mc}\,,
\eeq
is the Compton length associated to the constituent particle $m$. 

In the  case of galaxies, $m=m_{p}\cong 10^{-24}$gr, which is the proton
mass. $N$ is given by Eq.(\ref{10}) and we get
\beq
\label{13}
{\cal E}\simeq 10^{22}\mbox{cm}\simeq 10\mbox{kpc}\,,
\eeq
which is a typical scale of length for a normal galaxy. This fact means that 
the quantum parameter $\lambda_{c}$ and the number of constituents $N$
determine the astrophysical size ${\cal E}$ which is related to the stability
of the system. It is interesting to stress that this is the typical size 
where the rotation curve of a galaxy can be assumed flat \bib{binney} and,
in some sense, where the halo and the disk stabilize each other. If the 
characteristic time is given by Eq.(\ref{7}) and the geometrical scale is 
(\ref{13}), we get
\beq
\label{14}
v\simeq 10^{7}\mbox{cm/sec}\,,
\eeq
which is a typical rotational velocity for the outer components of a galaxy.
A further interesting quantity is the time after which the system can be 
considered virialized. It is
\beq
\label{15}
{\cal T}_{vir}\simeq 10\div 100{\cal T}_{rot}\simeq 1\div 10\mbox{Gyr}\simeq
10^{16\div 17}\mbox{sec}\,.
\eeq
Considering also the typical maximal extension of the halo which can be 
assumed
\beq
\label{16}
R\simeq 1\div 10{\cal E}\cong 10\div 100\mbox{kpc}\,,
\eeq
we get
\beq
\label{17}
v\simeq 10^{5\div 7}\mbox{cm/sec}\,,
\eeq
which is the range where are placed all the typical velocities of a normal 
galaxies, i.e. from the dispersion of velocities of stars 
($\sim 10^{5}$cm/sec) to the circular speed of a star in the disk
($\sim 10^{7}$cm/sec). We stress again that all these quantities are, in some
sense, related to the Planck constant $h$.

From a thermodynamical point of view, the fluctuative time (\ref{2}) can be 
defined as the ratio between the typical quantum mechanical size $h$ anf
the  typical Boltzmann ``size" $k_{B}T$, \ie
\beq
\label{18}
\tau\cong\frac{h}{k_{B}T}\,,
\eeq
then the temperature of the system is
\beq
\label{19}
T\cong\left(\frac{h}{k_{B}}\right)\frac{\sqrt{N}}{{\cal T}}\,.
\eeq
Let us note that Eq.(\ref{19}) can be rewritten in the form
\beq
\label{pippo}
(k_{B}T){\cal T}\cong h\sqrt{N}\,,
\eeq
which defines the thermal unit of emittance and also determines,
dividing  by $mc$, the
typical length range of interaction, Eq.(\ref{11}).

Let us now consider a galaxy as a thermodynamical system, then
\beq
\label{20}
\frac{1}{2}m_{p}\langle v^2\rangle=\frac{3}{2}k_{B}T\,.
\eeq
Using Eq.(\ref{19}), we get
\beq
\label{21}
\langle v^2\rangle\simeq\left(\frac{3h}{m_{p}}\right)
\frac{\sqrt{N}}{{\cal T}_{vir}}\,,
\eeq
and then the result (\ref{17}) is recovered for velocities.
Besides this ``thermal velocity'', it is straightforward to define a
``thermal wavelength''. Using the above results, we can write, in general,
\beq
\label{22}
\lambda_{T}\cong\frac{R}{\sqrt{N}}\,,
\eeq
and
\beq
\label{23}
\langle v\rangle\cong\frac{R}{{\cal T}}\cong\frac{\lambda_{T}}{\tau}\,,
\eeq
considering Eqs.(\ref{18}) and (\ref{19}), we get
\beq
\label{24}
\lambda_{T}\cong\frac{h}{\sqrt{m_{p}k_{B}T}}\,,
\eeq
which is of the order of atomic size as it must be for a proton.

In conclusion, we can say that the geometrical size, the kinematic and the 
stability of galaxies are strictly related to quantum mechanics. 
In other words, it seems that the structure of galaxies is ruled by quantum 
mechanics which prevents the onset of chaotic behaviour and, in some sense,
the dissipation of the constituents. Furthermore, it seems that the number of
constituents (the nucleons inside the stars) the geometrical global size $R$,
the classical law of force $F(R)$ have to combine in order to give a 
characteristic action of order $h$ to stabilize the systems.
A further step that the authors are going to face is to understand the
dynamics of such a feature and its connection to the cosmological evolution.

\vspace{2. mm}

{\bf Acknowledgements}\\
\noindent The authors would like 
to thank Luigi Galgani and Georgios Contopoulos for the enlightening
discussions and useful suggestions which allowed to 
 improve the content of the present paper.

\vspace{2. cm}

\begin{centerline}
{\bf REFERENCES}
\end{centerline}
\begin{enumerate}
\item\label{binney}
J. Binney and S. Tremaine, {\it Galactic Dynamics}
(Princeton University Press, Princeton, 1987).
\item\label{saasfee}
J. Binney, J. Kormendy, S.D.M. White, 
{\it Morphology and Dynamics of Galaxies}\\
12th Adv. Course of Swiss Society of Astr. and Astroph. Saas-Fee 1982\\
Eds. L. Martinet and M. Mayor, Geneva Observatory Editions.
\item\label{vorontsov}
B.A. Vorontsov--Vel'yaminov, {\it Extragalactic Astronomy}
Harwood Academic Pub. (London) 1987. 
\item\label{kolb}
E. W. Kolb and M. S. Turner, {\it The Early Universe}
(Addison--Wesley, New York, 1990).
\item\label{sakharov}
A. Sakharov, Zh. Eksp. Teor. Fiz. {\bf 49}, 245 (1965).\\
V. F. Mukhanov, H. A. Feldman and R. H. Brandenberger, 
Phys. Rep. {\bf 215}, 203 (1992).
\item\label{capozziello}
S. Capozziello, S. De Martino, S. De Siena and F. Illuminati\\
{\it Quantum Signature of Large Scale Cosmological Structures}\\
E-print gr-qcxxx (1998), submitted to Phys. Lett. {\bf A}. 
\item\label{lyndenbell}
D. Lynden--Bell, \mnras {\bf 136}, 101 (1967).\\
B.B. Kadomtsev and O.P. Pogutse, \prl {\bf 25}, 1155 (1970).
\item\label{galgani}
L. Galgani and A. Scotti, \prl {\bf 28}, 1173 (1972).
\item\label{bocchieri}
P. Bocchieri, A. Scotti, B. Bearzi, and A. Loinger
\pr {\bf 2A}, 2013 (1970).
\item\label{demartino}
S. De Martino, S. De Siena and F. Illuminati,
Mod. Phys. Lett. {\bf B 12}, 291 (1998).\\
S. De Martino, S. De Siena and F. Illuminati, {\it Inference 
of Planck action constant by a classical fluctuative
postulate holding for stable microscopic and macroscopic
dynamical systems}, E--preprint quant-phxxx (1998), 
submitted to Phys. Lett. {\bf A}.\\
N. Cufaro Petroni, S. De Martino, S. De Siena and F. Illuminati, 
{\it A stochastic model for the semiclassical collective
dynamics of charged beams in particle accelerators},
E--preprint physics/9803036 (1998), to appear in the
Proceedings of the International Workshop on
``Quantum Aspects of Beam Dynamics'', held in Stanford,
4--9 January 1998.
\item\label{calogero}
F. Calogero, Phys. Lett. {\bf A 228}, 335 (1997).
\item\label{stelle}
K. S. Stelle, Phys. Rev. {\bf D 16}, 953 (1977).\\
K. S. Stelle, Gen. Rel. Grav. {\bf 9}, 353 (1978).
\item\label{brans}
C. Brans and R. H. Dicke, Phys. Rev. {\bf 124}, 925 (1961). 
\item\label{bertolami}
O. Bertolami, J. M. Mourao and J. P\'erez--Mercader, 
Phys. Rev. {\bf B 311}, 27 (1993).
\item\label{contopoulos}
G. Contopoulos, private communication.
\item\label{fradkin}
E. S. Fradkin and A. A. Tseytlin, Nucl. Phys. {\bf B 201}, 469 (1982).
\item\label{sanders}
R. H. Sanders, Astron. Astrophys. Rev. {\bf 2}, 1 (1990).
\item\label{pioneer}
J.D. Anderson {\it et al.}, \prl {\bf 81}, 2858 (1998).
\item\label{avramidy}
E. G. Avramidy and A. O. Barvinky, Phys. Lett. {\bf B 159}, 269 (1985).

\end{enumerate}
 
\end{document}